\documentclass[sn-mathphys,Numbered]{sn-jnl}

\usepackage[T2A]{fontenc}                        
\usepackage[utf8]{inputenc}                       
\usepackage[english, russian]{babel}             
\usepackage[russian]{babel}                    
\usepackage{graphicx}

\usepackage{graphicx}%
\usepackage{multirow}%
\usepackage{amsmath,amssymb,amsfonts}%
\usepackage{amsthm}%
\usepackage{mathrsfs}%
\usepackage[title]{appendix}%
\usepackage{xcolor}%
\usepackage{textcomp}%
\usepackage{manyfoot}%
\usepackage{booktabs}%
\usepackage{algorithm}%
\usepackage{algorithmicx}%
\usepackage{algpseudocode}%
\usepackage{listings}%

\theoremstyle{thmstyleone}%
\theoremstyle{thmstyletwo}%

\theoremstyle{thmstylethree}%

\newcommand{\bb}{\begin{equation}}
\newcommand{\ee}{\end{equation}}

\newcommand{\bm}{\boldsymbol}

\begin{document}

\title{Electromagnetic radiation from a relativistic jet induced by a plane gravitational wave}



\author[1]
{\fnm{Vladimir} \sur{Epp}}\email{epp@tspu.ru}

\author*[1,2,3]{\fnm{Konstantin} \sur{Osetrin}}\email{osetrin@tspu.ru}

\author[1]
{\fnm{Elena} \sur{Osetrina}}\email{elena.osetrina@tspu.ru}

\affil[1]
{
\orgname{Tomsk State Pedagogical Unversity}, \orgaddress{\street{Kievskaya, 60}, \city{Tomsk}, \postcode{634061}, \country{Russia}}}

%
\affil[2]
{
\orgname{National Research Tomsk State University}, \orgaddress{\street{Lenina pr. 36}, \city{Tomsk}, \postcode{634050}, 
\country{Russia}}}

\affil[3]{
\orgname{Tomsk State University of Control Systems and Radioelectronics}, \orgaddress{\street{Lenina pr. 40}, \city{Tomsk}, \postcode{634050}, 
\country{Russia}}}

%


\abstract{
Electromagnetic radiation of a relativistic gas or plasma jet  in the field of a plane gravitational wave is investigated. The gravitational wave is considered as a  weak (linearized) field on flat Minkowski spacetime. It is assumed that the relativistic jet has large regions with uncompensated electric charge. The deformation of these areas under the action of a gravitational wave leads to the appearance of electric currents that generate electromagnetic radiation. The angular distribution of the intensity of this radiation is found. Cases are considered when the jet and the gravitational wave move in the same direction or towards each other.

}





\maketitle

\section{Introduction}
Direct detection of gravitational waves took place in the fall of 2015 \cite{Abbott_2016, Abbott_2019}. 
However, direct detection of those waves is only possible when they have a sufficiently short period and high enough intensity. For example, when black holes merge. Registration of long gravitational waves is quite problematic today. This is especially true for relic gravitational waves, which can have a fairly large wavelength  \cite{Odintsov,%
ODINTSOV2022100950,ODINTSOV2022136817,%
CAPOZZIELLO2021100867,NOJIRI2020100514,PhysRevD.98.024002,Osetrin356Universe_2023}.

Therefore, the proposal of alternative methods for detecting gravitational waves is very relevant.
Recently, many works have appeared devoted to the problem of interaction of gravitational waves with electromagnetic fields and charged particles.
As a result of such interaction, electromagnetic radiation can arise. In an early work, Heintzmann \cite{Heintzmann_1968} proposed a method of successive approximations for solving Maxwell's equations in the field of a spherical gravitational wave. Wickramasinghe \cite{Wickramasinghe_2015} and co-authors showed that charged particles can convert the energy of a gravitational wave into electromagnetic radiation.
Boughn \cite{Boughn_1975} solved Maxwell's equations for a point charge in the field of a plane gravitational wave by expanding the electromagnetic field potential in a series of spherical harmonics. An analysis of the coefficients of this expansion showed that the total radiation intensity, summed over harmonics, diverges. 
Sasaki and Sato \cite{Sasaki_1978} used the method of successive approximations to study the field of a relativistic point charge colliding with a plane gravitational wave. It was shown that the charge radiates into a narrow cone in the direction of its motion. At the same time, the intensity of the charge's radiation diverges in the direction of propagation of the gravitational wave. As we see, most authors encounter difficulties associated with divergences of various kinds when calculating the electromagnetic field of point charges interacting with a plane gravitational wave. The works cited above used a linearized theory of gravity. A number of works are devoted to the construction of  exact models of gravitational waves, including models of relic gravitational waves \cite{Osetrin2022894,Osetrin325205JPA_2023,Osetrin1455Sym_2023,%
ObukhovUniverse8040245,ObukhovSym15030648}

 Radiation of a stationary charged cloud  under the action of a gravitational wave is studied in \cite{Epp:2023jeg}. The frequency of this radiation coincides with the frequency of the gravitational wave, which under normal conditions is considered very small. One of the main sources of gravitational waves are close pairs of stars or black holes. The minimum observable period of rotation of such pairs is about half an hour (AM Canum Venaticorum stars). Registration of radio waves of such frequency is practically impossible. However in the case of a plasma or gas jet with an unevenly distributed charge, the Doppler effect can lead to a significant increase in the frequency of radiation induced by a gravitational wave.
 
 In this paper we investigate electromagnetic radiation induced by a gravitational wave in a relativistic  jet of plasma or gas. We assume that instabilities in such a jet can generate significant  regions of uncompensated electric charge. The gravitational wave deforms these regions in a known manner, which leads to displacement of charges and the appearance of electric currents. This process generates a variable electromagnetic field and, in particular, electromagnetic radiation.


\section{A model of relativistic jet}
We consider a model of a relativistic jet of gas, dust or plasma in which there are fairly large regions with a predominance of positive or negative charge. A plane gravitational wave incident on the jet, and can excite the relative motion of charges and the currents associated with this motion. Currents, in turn, generate an electromagnetic field, including a radiation field.
The properties of this radiation depend on the speed of particles in the jet, the charge distribution by volume, the size and shape of the charged region.

To obtain analytical expressions for the analysis of radiation properties, we simplify the model of such a region as follows. The charged region has a cylindrical shape of radius $r_0$ and length $L$, the charge density within this region is constant and equal to $\rho$. 
We will assume that the relativistic jet containing charged regions is in the field of a weak plane gravitational wave with metric
\begin{align}
g_{\mu\nu}&=\eta_{\mu\nu}+h_{\mu\nu},\quad h\ll 1,
\label{g}
\\
h_{\mu\nu}&=a_{\mu\nu}\exp(i\kappa_\sigma x^\sigma),
\end{align}
where $\eta_{\mu\nu}$
is the metric tensor of Minkowski space, $a_{\mu\nu}$ is the amplitude of gravitational waves, $\kappa^\sigma=(\omega/c, 0,0,\omega/c)$ is the wave vector of gravitational waves, $\omega$ is its frequency. In in case of transverse traceless gauge, the wave amplitude can be represented as
\bb\label{matr}
 a_{\mu\nu}=
 \begin{bmatrix}
    0       &0 &0 & 0 \\
   0  & a & b &0\\
    0&b&-a&0 \\
   0& 0 & 0 &0
\end{bmatrix}
 \ee
for a wave propagating in the direction of the $z$ axis. To simplify the calculations and concentrate on the physical effects, we set $b=0$.

\begin{figure}[ht]\center
\includegraphics[width=5.6cm]{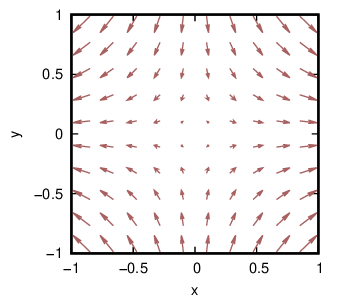}
\caption{Currents induced by gravitational wave in a charged cloud}
\label{(1)}
\end{figure}

To calculate the electromagnetic radiation induced by a gravitational wave, we proceed as follows. We find the induced electromagnetic field in the reference frame associated with the jet, then we move to the reference frame associated with the observer. In the article \cite{Epp:2023jeg} the radiation field of a stationary charged cloud, in the field of a gravitational wave is found. In particular, it is shown that the gravitational wave excites an electric current in the cloud in a plane perpendicular to the direction of propagation of the wave with a current density (from here onward $x^0=ct,\, x^1=x, \,x^2=y, \, x^3=z$)
 \bb
\bm j(x,y,z)=
\frac{1}{2} \left(x,-y,0\right)\rho a \omega\sin \kappa(ct-z).
\label{31}
\ee 
From the continuity equation it follows that the charge density in the cloud remains constant. The current density map (\ref{31}) for a certain fixed moment of time is shown in Fig. \ref{(1)}

\section{Radiation from a charged cloud}
In our paper \cite{Epp:2023jeg} the angular distribution of the radiation intensity of a charged cylindrical cloud under the action of a gravitational wave with the metric (\ref{g})-(\ref{matr}) is obtained
\bb
\frac{d I}{d\Omega}=
\frac{\pi  a^2 \rho^2 \omega^4 R^6 L^2}{8c^3}\frac{\sin^2\chi}{\chi^2} \frac{1}{u^2}J_2^2(u)
\label{111}.
\ee
Here $d I/d\Omega$ is the energy emitted into a solid angle $d\Omega$ per unit time, $a$ is the amplitude of the gravitational wave, $\rho$ is the charge density in the cloud, $\omega$ is the frequency of the gravitational wave, $R$ and $L$ are the radius and length of the cloud, respectively,
\bb\label{chiu}
\chi=\frac{L\omega}{2c}(1-n\cos\theta), \quad u=\frac{\omega n R }{c}\sin\theta,
\ee
$n$ is the refractive index of the cloud matter, $J_2(u)$ is the Bessel function, $\theta$ is the angle between the direction of radiation and the wave vector of the gravitational wave. The frequency of electromagnetic radiation is equal to the frequency of the gravitational wave. 

The angular distribution of radiation is determined by the function
\bb\label{ft}
f(\theta)=
\left(\frac{\sin\chi}{\chi} \frac{J_2(u)}{u}\right)^2.
\ee
If the cloud size is measured in gravitational wavelengths, 
\[
l_\perp=\frac{\omega R}{2\pi c},\quad l_\parallel =\frac{\omega L}{2\pi c},
\]
then the variables in the equation (\ref{ft}) are equal to
\bb
\chi=\pi l_\parallel (1-n\cos\theta), \quad u=2\pi n l_\perp\sin\theta,
\ee

\begin{figure}[ht]\center
\includegraphics[width=5.6cm]{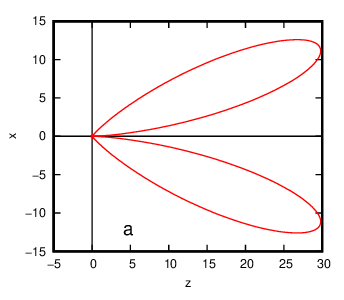}\quad \includegraphics[width=5.6cm]{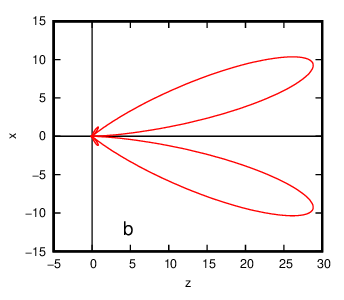}
\includegraphics[width=5.6cm]{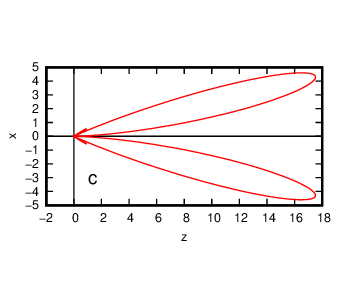}\quad \includegraphics[width=5.6cm]{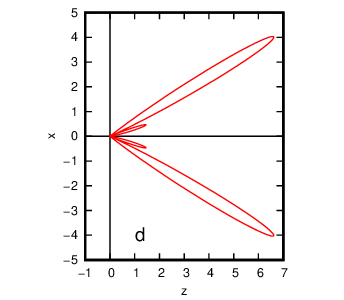}
\caption{Angular distribution  $f(\theta)\cdot 10^3$ of electromagnrtic radiation. $a$:  $l_\perp=l_\parallel=1$, $n=1$; $b$ --  $l_\perp=l_\parallel=1$, $n=1.2$; $c$ -- $l_\perp=1,\,l_\parallel=10$, $n=1$; $d$ -- $l_\perp=1,\,l_\parallel=10$, $n=1.2$.}
\label{(2)}
\end{figure}

The picture of the angular distribution of electromagnetic radiation is presented in Fig. \ref{(2)} for different values
of the reduced dimensions of the cloud and the refractive index of the cloud substance.
The plots show that the main part of the radiation is concentrated within the cone forming an acute angle with the direction of propagation of the gravitational wave. Outside and inside the main cone of radiation there are side lobs of radiation. Their intensity and number depend on the size of the charged cloud and on the permittivity of the cloud substance. The angular width of the main lobe depends significantly on the size of the cloud in units of gravitational wave lengths. This is due to the fact that the currents in the cloud are induced by the gravitational wave and, therefore, change coherently with time.

We use Eq. (\ref{111}) to calculate the electromagnetic radiation of a relativistic  jet. According to our model, such a jet may contain regions with distributed uncompensated charge. In the reference frame associated with such a moving region, Eq. (\ref{111}) is applicable. All quantities in this associated reference frame will be denoted by primes:
\bb
\frac{d I'}{d\Omega'}=
\frac{\pi  a'^2 \rho'^2 \omega'^4 R'^6 L'^2}{8c^3}\frac{\sin^2\chi'}{\chi'^2} \frac{J_2^2(u')}{u'^2}
\label{111'}.
\ee
We will assume that the jet moves parallel to the $z$ with the velocity  $v_z$ which can be positive or negative. If the jet moves in the direction of the gravitational wave, $v_z>0$, if the  velocity is directed opposite to the gravitational wave, $v_z<0$.

The Lorentz transformations into the observer's frame of reference for the quantities  in the Eq. (\ref{111'}) have the form
 \cite{Landau_II}
\[
\cos\theta'=\frac{\cos\theta-\beta_z}{1-\beta_z\cos\theta},\quad \sin\theta'=\frac{\sqrt{1-\beta_z^2}\sin\theta}{(1-\beta_z \cos\theta)},\quad \omega'=\omega\sqrt{\frac{1-\beta_z}{1+\beta_z}}
\]
\[
d\Omega'=\frac{(1-\beta_z^2)d\Omega}{(1-\beta_z \cos\theta)^2},\quad \frac{d I'}{d\Omega'}=\frac{(1-\beta_z \cos\theta)^3}{(1-\beta_z^2)^2}\frac{d I}{d\Omega}
\]
The Lorentz transformations do not change the transverse component of the gravitational wave amplitude: $a'=a$. The variables (\ref{chiu}) are transformed as follows
\begin{align}\label{chiu'}
\chi'=&\frac{L'\omega'}{2c}(1-n'\cos\theta')=\frac{L\omega}{2c(1+\beta_z)}\left(1-n'\frac{\cos\theta-\beta_z}{1-\beta_z\cos\theta}\right),\\ 
u'=&\frac{\omega' n' R' }{c}\sin\theta'=\frac{\omega Rn' }{c}\frac{(1-\beta_z)\sin\theta}{1-\beta_z \cos\theta}.
\end{align}
We leave the values of permittivity and refractive index in the accompanying frame of reference, 
since they depend on the internal parameters of the substance in the jet.

\begin{figure}[ht]\center
\includegraphics[width=5.8cm]{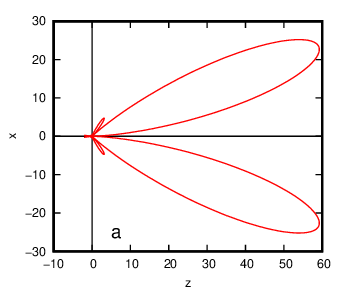}\quad \includegraphics[width=6cm]{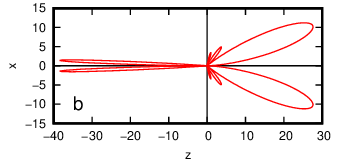}
\includegraphics[width=5.8cm]{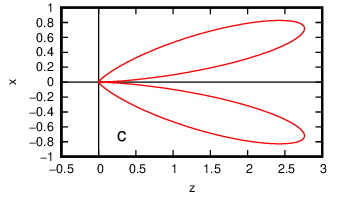}\quad \includegraphics[width=6.2cm]{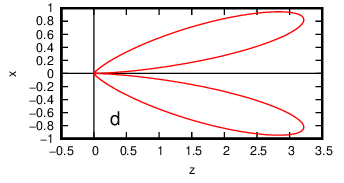}
\caption{Angular distribution  $f_{rel}(\theta)\cdot 10^3$ of electromagnrtic radiation from relativistic cloud. All plots are built for $l_\perp=l_\parallel=1$, but different $n$ and $\beta_z$. From top left to bottom right: $a$:  $n=1,\,\beta_z=-0.8$; $b$: $n=1.1,\,\beta_z=-0.8$; $c$: $n=1,\,\beta_z=0.8$; $d$: $n=1.1,\,\beta_z=0.8$.}
\label{(3)}
\end{figure}

After these transformations, the angular distribution of radiation in the observer's frame of reference takes the form
\bb
\frac{d I}{d\Omega}=
\frac{\pi  a^2 \rho^2 \omega^4 R^6 L^2(1-\beta_z)^4}{8c^3(1-\beta_z \cos\theta)^3}\frac{\sin^2\chi'}{\chi'^2} \frac{J_2^2(u')}{u'^2}
\label{111'}.
\ee
Fig. \ref{(3)} shows the angular distribution of radiation from a charged cloud moving at a speed of $\beta_z=\pm 0.8$ towards the gravitational wave (inserts $a,\,b$, the gravitational wave propagates to the right, the cloud moves to the left) and in the direction of propagation of the gravitational wave (inserts $c,\,d$, both, the gravitational wave and the cloud move to the right).
The radiation patterns are plotted according to the function
\bb
f_{rel}(\theta)=
\frac{(1-\beta_z)^4}{(1-\beta_z \cos\theta)^3}\frac{\sin^2\chi'}{\chi'^2} \frac{J_2^2(u')}{u'^2}.
\label{111'}.
\ee

If we compare Fig. \ref{(2)}  and Fig. \ref{(3)}, we can see that the radiation is amplified in the direction of the cloud's motion in accordance with the Doppler effect. As already noted, the main part of the radiation is emitted in the same direction as the  gravitational wave propagates. If the cloud moves toward the gravitational wave, the back lobes are amplified significantly. This is  clearly seen in Fig. \ref{(3)}b.

If the cloud and the gravitational wave move in the same direction, then the electromagnetic radiation relative to a stationary observer is significantly weakened. The relativistic motion of the charged cloud leads to the amplification of only the side and back lobes of the radiation pattern.  In this case,  motion of the cloud  does not significantly improve the conditions for observing induced radiation.

\section{Discussion}
The main advantage of a relativistic cloud in terms of recording electromagnetic radiation is the Doppler frequency shift. A stationary cloud radiates with the same frequency as the metric tensor of a gravitational wave changes. The period of gravitational waves generated by close binary cosmic objects is days or hours at best. 
Registration of radio waves of such length is associated with known technical difficulties. The relativistic motion of the charged cloud towards the Earth significantly reduces the wavelength of the induced electromagnetic radiation. 

The frequency of radiation in the reference frame accompanying the cloud is determined by 
\bb
\omega'=\omega\sqrt{\frac{1-\beta_z}{1+\beta_z}},
\ee
where $\omega$ is the frequency of the gravitational wave in the frame of reference of a stationary observer. The frequency of radiation relative to the stationary frame of reference is equal to
\bb
\omega_{rad}=\omega'\frac{\sqrt{1-\beta_z^2}}{1-\beta_z\cos\theta}=\frac{\omega(1-\beta_z)}{1-\beta_z\cos\theta}.
\ee
If the cloud is moving towards the gravitational wave ($\beta_z<0$) and the direction to the observer makes a small angle with the direction of the velocity ($\theta\sim\pi$), then $\omega_{rad}\sim\gamma^2$ where $\gamma=(1-\beta_z^2)^{-1/2}$ is the relativistic factor. Relativistic jets accompanying gamma-ray bursts can have a speed  up to $\gamma\sim 10^3$ \cite{Dereli_2022}.

In addition, binary stars rotating in very elongated orbits generate gravitational waves of a wide spectrum. The short-wave part of this spectrum, combined with the Doppler effect of frequency increase, can induce electromagnetic radiation in relativistic jets with a sufficiently short wavelength, accessible for registration by modern means.

\section*{Acknowledgments}
The study was supported by the Russian Science Foundation, grant \mbox{No. 23-22-00343},
\\
\url{https://rscf.ru/en/project/23-22-00343/}.



\end{document}